# Design of Run time Architectures for Real time UML Models an Actor Centric Approach

[1]PVRR Bhogendra Rao, [2]V Kamakshi Prasad
[1]Scientist, DRDL, Hyderabad, rao_punyamurty@yahoo.com
[2]Professor, JNT University, Hyderabad

## Abstract

**Although a lot of research has taken place in Object Oriented Design of software for Real Time systems and mapping of design models to implementation models, these methodologies are applicable to systems which are less complex and small in source code size. However, in practice, the size of the software for real time applications is growing. The run time architecture of real time applications is becoming increasingly complex. In this paper, we present a generic approach for mapping the design models to run time architectures resulting in combination of processes and threads. This method is applied in development of a communication subsystem of C4I complex and shall be presented as a case study.**

## 1. INTRODUCTION

The real-time systems are becoming increasingly complex and large in terms of source code size. This is particularly the case with C4I application domain. C4I applications are safety-, mission- and time-critical in nature. The increasing complexity and the sophisticated demands of such systems in terms of safety, reliability, and performance require the use of rigorous development methodologies and CASE tools for reliable software development.

Object-Oriented modeling and design has become the most preferred methodology to the software designers, ever since its advent, for handling the complexity of the software. In the recent past it has become popular in the real-time domain also and a number of efforts have been put to apply this methodology towards modeling and design of real-time systems to gain the above advantage. Object oriented analysis and design models are comprised of various artifacts and concepts such as classes, objects and state charts. By using modeling abstractions that are closer to the problem space as well as visual notations, object-oriented modeling and design facilitates the design process and promotes a better understating of the design. Furthermore, such models also facilitate various forms of analysis and simulation to help in the design process.

The Object-oriented methodology has become more attractive to the real-time application developers with its support for real-time application development in UML 2.0 in the form of added artifacts such as capsules, ports, protocols and timing diagrams. A number of CASE tools are commercially available which support UML 2.0. Popular among them are Rose Real-time from IBM, Tau from Telelogic, Rhapsody from i-Logix. These CASE tools support not only visual modeling but also automatic code generation. However, the performance of the real-time application is highly dependent on its run-time architecture. The CASE tools, which support UML 2.0, neither do support any automatic mapping nor do provide any guidelines to the designer in identification of run-time architecture.

Scenario based techniques have been proposed in [3] for automatic mapping of the design models to implementation models. However, there are a number of disadvantages with these approaches. (1) In practice, a large real-time system such as C4I is implemented as a combination of processes and threads, rather than as a single process containing a number of threads. (2) If the entire system is implemented as a single process, it would not be fault tolerant. In case, if any one of the threads abnormally terminates due to an error, the entire process would terminate causing a total system failure. (3) The approach is based on the scenarios, which are nothing but the functions that the system should perform. In a sense, it is a procedure oriented approach rather than object-oriented approach.

In a use case driven development, these use cases are ultimately translated in to a set of classes that can be implemented. However, the standard use case driven development process leads to static architecture of the software in terms of classes and relationships among these classes.

In this paper we present a use case driven approach for evolving run-time architecture. We are presenting an approach for mapping the UML components onto the runtime architecture, which is applicable to large real-time systems and this approach results in a combination of processes and threads. This approach yields scalable run-time architecture with graceful degradation for the software. We would explain the approach with a communication application from C4I domain.

## 2. OVERVIEW OF UML 2.0

## 2.1 *UML* 2.0

The UML includes the artifacts required for real-time system development from UML 2.0. These artifacts are imported from the ROOM (Real-time Object Oriented Methodology) methodology. The ROOM method adopts an operational approach to system analysis, design and implementation. It is based on establishing early operational models of the system and then refining them to implementation. It uses the concept of executable models which evolve from requirements to design to implementation. A UML 2.0 / ROOM executable model is a set of coherent structure and behavior view which can be compiled and executed on a variety of simulation and/or target platforms.

Modeling of real-time systems with UML 2.0 is performed by designing *active classes* or Capsules, which are encapsulated, concurrent objects communicating via point-to-point links. Inter actor communication is performed exclusively by sending and receiving *messages* via interface objects called *ports*. A message is a tuple consisting of a signal name, a message body (i.e., data associated with the message), and an associated message priority.

The behavior of an actor is represented by an extended state machine called a *State diagram*, based on the statechart formalism [12]. Each actor remains dormant until an *event* occurs, i.e., when a *message* is received by an actor. Incoming messages trigger transitions associated with the actor's finite state machine. Actions may be associated with transitions, as well as entry and exit points of a state. The sending of messages to other actors is initiated by an action. The finite state machine behavior model imposes that only one transition at a time can be executed by each actor. As a consequence, a *run-to-completion* paradigm applies to state transitions. This implies that the processing of a message cannot be preempted by the arrival of a new (higher priority) message for the same actor. However, as explained later in a multi-threaded or hybrid implementation, the processing may be preempted by other higher priority threads.

UML 2.0 / ROOM support the notion of a *composite* state, which can be decomposed into sub-states. composition of a state into sub-states can be taken upto any arbitrary level in a recursive manner. The current state of such a system is defined by a nested chain of states called a *state context*. The behavior is said to be simultaneously "in" al of these states. Transitions on the innermost current state take precedence over equivalent transitions in higher scopes. As event for which no transition is triggered at all levels of the state hierarchy is discarded unless it is explicitly deferred.

## 2.2 Communication Network Interface for C4I: A Case Study

We use a communication system of a C4I complex to illustrate the concepts explained in this paper.

Typically a C4I complex is comprised of a number of interacting elements and these elements are connected through a communication subnet. The communication subnet is a vital element for the success of the mission in a network centric warfare scenario. Similarly the network plays an important role in the maintenance of the C4I software. Changes in the communication network due to technological advances and/or user requirements, cause changes in the C4I software. These changes may be very expensive in terms of reliability of the software. So, to isolate the C4I system from changes in the communication system, most of the C4I complexes use another subsystem to handle the real-time communication required for the mission as shown in the Fig 1.

The major functions of the communication interface system are as follows:

1. Packetization and reassembly
2. Protocol conversion
3. Fault tolerance at media level
4. Session maintenance
5. Priority based on the type of the data
6. Security and authentication of data
7. Logging
8. Monitoring health of communication equipment
9. On-line status display and alerts

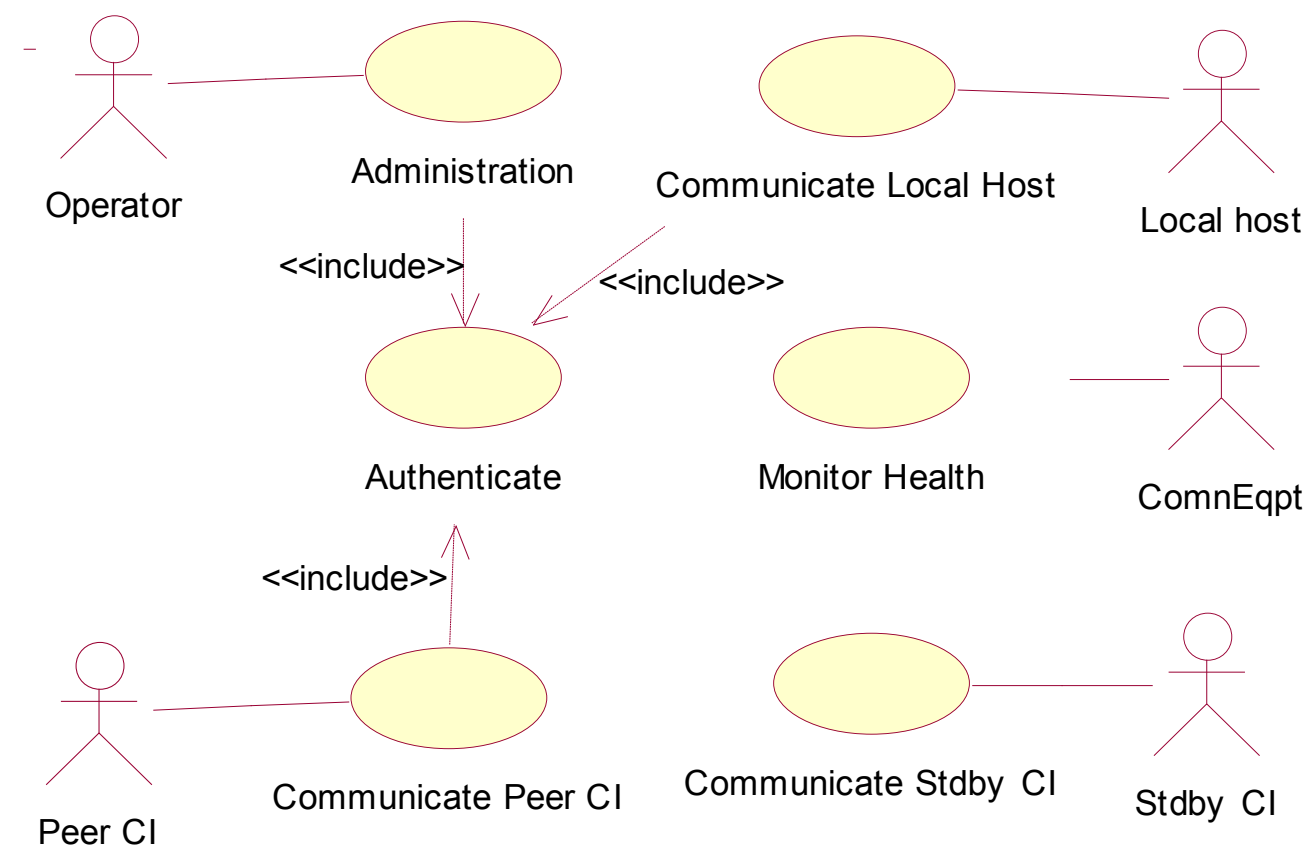

Figure 1. Use Case Diagram for Communication Interface System

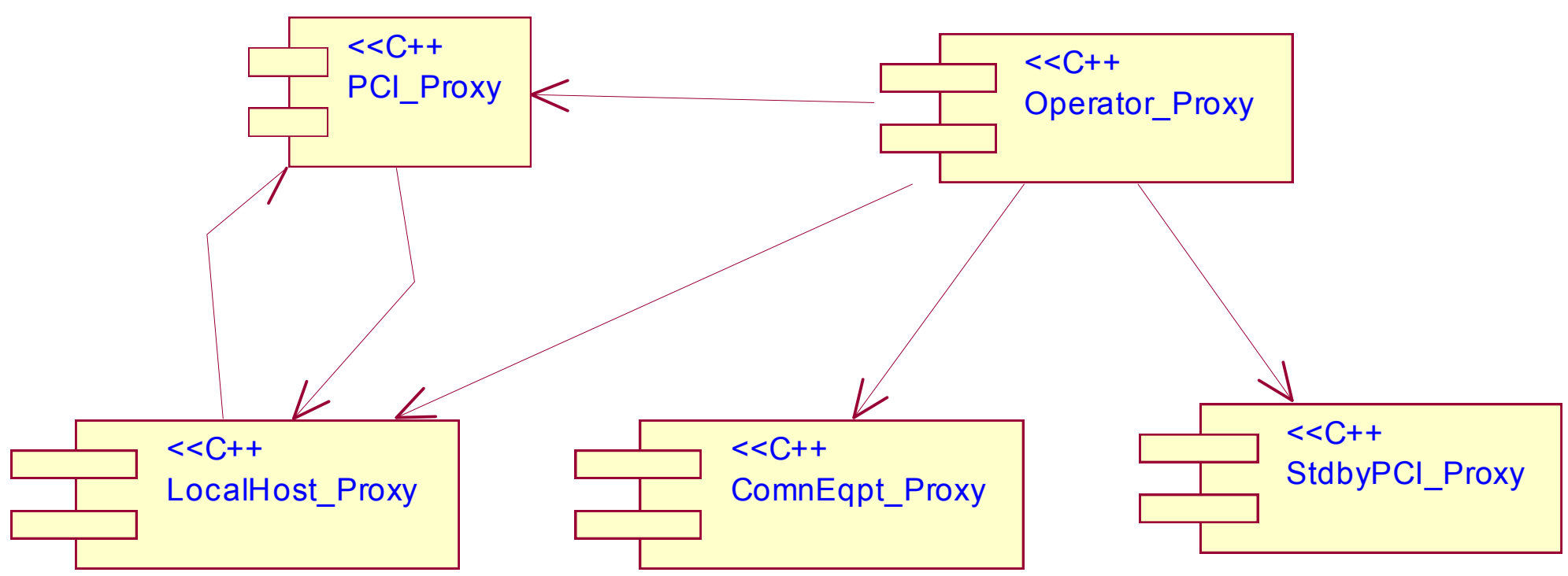

Figure 2. Component Diagram

As shown in the Use Case diagram of Fig. 1 the communication Interface needs to communicate with a number of peer interfaces, stand by interface (needed to take-over the functionality in the event of failure of main), a number of communication equipment, and a number of local hosts.

The communication Interface receives data in the form of UDP messages from the local hosts over Ethernet and breaks these messages into smaller packets before sending these packets to the suitable peer communication interface through appropriate communication link after conversion into required protocol. The peer communication server receives these packets, assembles the messages, and hands them over to the local host. Optionally security and authentication are also involved in this process. These numbers of peer interfaces, communication equipment, local hosts may vary for different element of C4I complex. So, there can be multiple instances of the same actor. The run-time architecture should facilitate the scalability of the system to different number of actors.

## 3. THE GENERIC MAPPING APPROACH

The UML supports use case driven development of the software, which the software engineers recognized as the natural approach to translate the requirements to design models to implementation models. These requirements are modeled as Use cases in UML and external actors trigger these Use cases. It is required to model the behavior of the system in response to the events generated by these external actors in the design of software for real time systems. So, we found it is natural to design the architecture of the software in an **actor centric approach**.

This approach encompasses the following steps:

1. Model the external and internal use cases.
2. Model processes
3. Model IPC mechanisms.
4. Model the concurrency within the process.

### 3.1 Model External and Internal Use cases

This is the first step in object oriented analysis and design of any software system. This process involves identification the external systems with which the system under realization is going to interact with. Some of these external systems may generate events and some may receive. Each of these interacting systems will be an actor in our use case model. Now, identify the functions of the software system as seen by each actor. Model each end to end sequence of actions as including variants that the system can perform for each actor as one Use case [1]. Document the behavior of each use case. Identify the common functions among the use cases and model them as either **include**d use cases or **extend**ing use cases as the case may be. For details on use cases refer [2].

### 3.2 Model Processes

Group the use cases with respect to the actors that trigger them. A use case can be present in multiple groups, as it can be triggered by multiple actors. Each of these groups gives the view of the software system as seen by that corresponding actor; hence we call each group a **View Case**. Map each View Case onto one **POSIX process**.

Let $U$ be the set of use cases representing the requirements that the software system must meet.

$$U = \{u_1, u_2, u_3, \ldots\ldots\ldots u_n\}$$

Now, these requirements are divided into subsets – not necessarily mutually disjoint – each subset corresponding one actor. Each subset is a collection of all the requirements that the system should meet with respect to one actor. As explained above each of these subsets represents the system as viewed by a particular actor. Let $V$ be the set of views. The set $V$ represents the runtime view of the software whereas $U$ represents the static view of the software.

$$V = \{v_1, v_2, v_3, \ldots\ldots\ldots v_n\}$$

Each of these subsets $v_i$ is realized as a UML 2.0 run-time component and corresponds to a POSIX process.

The following issues arise in this phase

1. How to model when one use case being triggered by multiple actors
2. How to map when the multiplicity of a particular actor is more than one.
3. How to map **Include**d and **extend**ing use cases

All of the above cases arise with reusable modules. These situations can be addressed in two different ways depending on the fault tolerance requirements and available physical memory.

When fault tolerance is of prime importance, create one process for each actor / actor instance for cases 1 and 2 to provide the services required by that actor. Each process interacts with one actor / actor instance alone and responds to the events of that actor. When the size of the application software is limited by the available physical memory, create one process for all the actors that trigger a use case to get a specific service from the software system (case 1) and one process for all the instances of an actor class (case 2).

Case 3 is more common in modeling the requirements. The common functions among the use cases are modeled as **include**d or **extend**ing use cases to avoid describing same flow of events several times. Organizing use cases by extracting common behavior (through include relationships), and distinguishing variants (through extend relationships) is an important part of creating a simple, balanced, and understandable set of use cases for a system[2]. The choice of mapping the **include**d or **extend**ing use case not only depends on fault tolerance requirements and physical memory limitations, but also on the size of the code generated to realize these use cases. If the size of the application software is limited by the size of physical memory and/or the size of the code generated to realize the **include**d / **extend**ing is large it is suggested to map a set of such use cases onto a process. But, this process may become a single point failure in the system and may be concern from fault tolerance point of view.

The other choice is to statically integrate the code for realizing the **include**d or **extend**ing use at the including or extending points in the other use cases. This option causes increase in the size of the processes.

In the realization of the software described in the case study, *authenticate* use case is included in three other use cases. The code to realize this use case is statically integrated with the other use cases, because the size of the code is less and the fault tolerance requirements are high.

In the *Communication Interface* application, the following processes are realized.

1. Operator's Interface (GUI)
2. Local Host Interface
3. Standby CI Interface
4. Monitor Health
5. Peer CI Interface

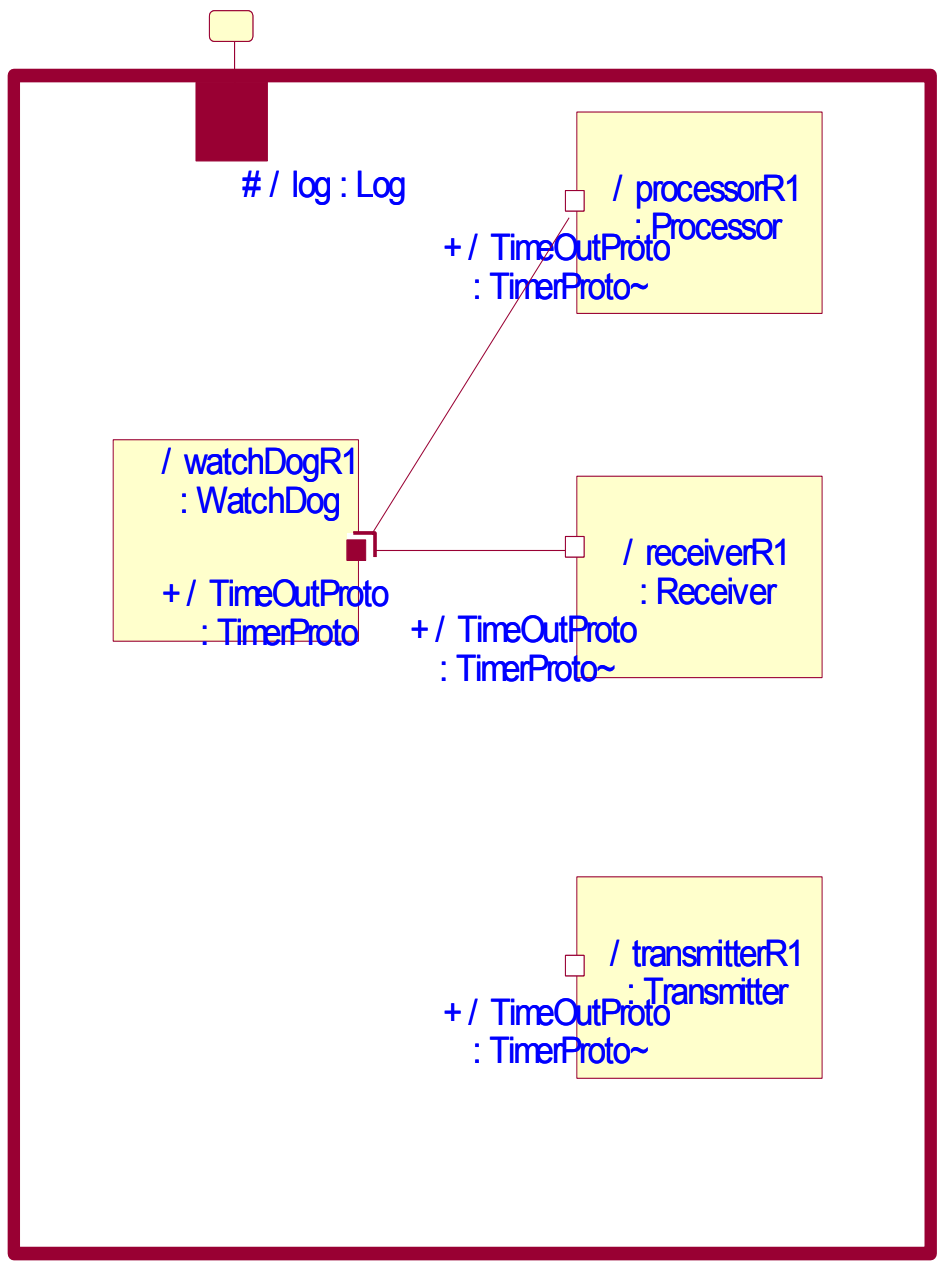

Figure 3. Structure diagram

Out of these, the *local Host Interface* and *Peer CI Interface* processes will be multiple instances – as many as local hosts and peer CI systems. In the final version, there are 2 processes for each of the local hosts and 6 Peer CI processes for each of the Peer CI's.

### 3.3 Model IPC Mechanisms

Each component acts as a proxy for the corresponding actor. As a result of an event generated by an actor, *A*, a process may generate inputs for another actor, *B*. However, in this actor centric approach, another process acts as a proxy for the actor, *B*. So, there is a need for IPC mechanism between the proxies of actors, *A* and *B*. In order to design an effective IPC mechanism, the first step is to understand the dependency among different processes. We have used component diagram to understand the dependency among the processes as shown in fig. 2. The IPC mechanism is highly application specific and general guidelines are available in the literature for choosing suitable IPC mechanism. In the current application, we have used both POSIX shared memory and POSIX message queues.

The shared memory IPC is used for communicating periodic information, such as health status and for communication from one process to multiple processes. The message queue IPC mechanism is used for communicating asynchronous information and for point-to-point communication. Thus, for communication from Operator's Interface process to other processes the IPC mechanism was shared memory. Between Peer CI Interface and Local Host Interface process the IPC mechanism was message queue.

The IPC mechanisms are wrapped in passive classes in the implementation of Communication Interface application.

### 3.4 Model Concurrency

Each process realizes one view – which is a set of more than one use case – and may interact with more than one actor (as a result of limitation of available physical memory). In such a situation, each process may have to concurrently process multiple events – each event triggering a scenario of the use case. Saksena, *et al* have proposed a scenario based multi threading approach in [3] for automated implementation of executable object oriented models for real-time embedded control systems. This approach is more suitable for reactive systems and each thread encompasses an end-to-end sequence of actions of scenario. A number of papers were found in the literature with similar proposals.

However, in the realization of the current application we have followed fixed priority periodic task model, as this model was found reliable in the literature because of its ready proof of schedulability. Fig 3 shows a generic structure of each process. Each process has at least four threads – a watch dog timer, a thread for receiving data from other processes, a thread for processing the received data and thread for dispatching the result to other processes. The business logic is built into the *processor* thread. The *receiver* and *transmitter* contain the objects of required IPC mechanism.

## 4. CONCLUSIONS

We have presented our experience of developing a large real-time application using UML 2.0. This software is realized using IBM's Rose Real-Time CASE tool on SUN Solaris operating system. One of the advantages of using a CASE tool is the model can be developed independent of the final target platform. The model can be developed, compiled and validated on a host environment and can be deployed on the target at a later date after validation. The code can be automatically generated by CASE tool for different targets.

The model proposed in this paper results in software that gives the required graceful degradation for real-time mission critical applications. In the current implementation we have used different process instances to communicate with different instances of Peer CI actor. Even if a process terminates abnormally due to an error, the communication link is lost with only the corresponding Peer CI. If this software were designed purely based on multi threading, even if one thread ab-normally terminates due to error, entire application would terminate causing a total failure of the system.

This approach results in software that is highly scalable. Though the Communication Interface is designed and deployed to communication with 2 local hosts and 6 Peer CI's, it can be easily configured to communicate with more local hosts or Peer CI's by creating new process instances to communicate with the new actors, without redesigning the software.

## BIOGRAPHY

**PVRR Bhogendra Rao** is scientist in Defense R & D Organization (DRDO), Hyderabad. Currently he is the research student of JNTU, Hyderabad.

**Dr. Kamakshi Prasad** acquired his doctorate from Indian Institute of Technology, Chennai, India. Currently he is professor in Jawaharlal Nehru Technological University, Hyderabad, India. His research interest areas are Computer Networks, Software Engineering and Image Processing etc.